\documentclass[12pt, onecolumn, draftcls]{IEEETran}
\usepackage{epsfig, psfrag, amsmath, amssymb, amsfonts, latexsym, eucal}%, endfloat}
\newtheorem{thm}{Theorem}%[section]

\newtheorem{lem}{Lemma}

\newtheorem{defn}{Definition}

\begin{document}

%%%%%%%%%%%%%%%%%%%%%%%%%%%%%%%%%%%%%%%%%%%%%%%%%%%%%%%%%%%%%%%%%%%%%%%%%%%%%%%%%%%%%%%%%%%%%%%%%%%%%%%%%%%%
\title{On Minimax Robust Detection of Stationary Gaussian Signals in White Gaussian Noise}
\author{Wenyi Zhang, {\it Member, IEEE}, and H.~Vincent Poor, {\it Fellow, IEEE}
\thanks{The work of W. Zhang was supported in part by Program for New Century Excellent Talents in University (NCET), and the Fundamental Research Funds for the Central Universities. The work of H.~V. Poor was supported in part by the U.S. Office of Naval Research under Grant N00014-09-1-0342.}
\thanks{W. Zhang is with Department of Electronic Engineering and Information Science, University of Science and Technology of China, Hefei 230027, China. Email: {\tt wenyizha@ustc.edu.cn}. H.~V. Poor is with Department of Electrical Engineering, Princeton University, Princeton, NJ, USA. Email: {\tt poor@princeton.edu}.}
}

\maketitle

%%%%%%%%%%%%%%%%%%%%%%%%%%%%%%%%%%%%%%%%%%%%%%%%%%%%%%%%%%%%%%%%%%%%%%%%%%%%%%%%%%%%%%%%%%%%%%%%%%%%%%%%%%%%
\begin{abstract}
The problem of detecting a wide-sense stationary Gaussian signal process embedded in white Gaussian noise, where the power spectral density of the signal process exhibits uncertainty, is investigated. The performance of minimax robust detection is characterized by the exponential decay rate of the miss probability under a Neyman-Pearson criterion with a fixed false alarm probability, as the length of the observation interval grows without bound. A dominance condition is identified for the uncertainty set of spectral density functions, and it is established that, under the dominance condition, the resulting minimax problem possesses a saddle point, which is achievable by the likelihood ratio tests matched to a so-called dominated power spectral density in the uncertainty set. No convexity condition on the uncertainty set is required to establish this result.
\end{abstract}
%%%%%%%%%%%%%%%%%%%%%%%%%%%%%%%%%%%%%%%%%%%%%%%%%%%%%%%%%%%%%%%%%%%%%%%%%%%%%%%%%%%%%%%%%%%%%%%%%%%%%%%%%%%%

%%%%%%%%%%%%%%%%%%%%%%%%%%%%%%%%%%%%%%%%%%%%%%%%%%%%%%%%%%%%%%%%%%%%%%%%%%%%%%%%%%%%%%%%%%%%%%%%%%%%%%%%%%%%
\section{Introduction}
\label{sec:intro}

Many signal detection problems can be modeled by the following hypothesis testing problem:
\begin{eqnarray}
\label{eqn:problem}
&&\mathcal{H}_0:\quad Y_n = N_n,\quad n = 0, 1, \ldots, N - 1,\nonumber\\
&&\mathcal{H}_1:\quad Y_n = S_n + N_n,\quad n = 0, 1, \ldots, N - 1,
\end{eqnarray}
where $N$ denotes the length of the observation interval. The noise samples $\{N_n\}$ are independent and identically distributed (i.i.d.) Gaussian random variables with zero means and variances $\sigma^2$, {\it i.e.}, $N_n \sim \mathcal{N}(0, \sigma^2)$. The stochastic signal, $\{S_n\}$, is a wide-sense stationary (WSS) Gaussian process with mean zero and power spectral density (PSD) $\phi(\omega)$, $\omega \in [-\pi, \pi]$.

For each $N$, the hypothesis testing problem is between two $N$-dimensional zero-mean Gaussian distributions, and we shall denote a detector by $\delta_N(\cdot)$, which maps the $N$-dimensional observation $\underline{y} = [y_0, y_1, \ldots, y_{N-1}]^T$ into $\{\mathcal{H}_0, \mathcal{H}_1\}$. As $N$ grows without bound, the detectors $\delta_1, \delta_2, \ldots$ constitute an infinite sequence, denoted by $\underline{\delta}$.

To characterize the discrimination capability for a specific detector sequence, a convenient performance metric is the exponential decay rate of the miss probability ({\it i.e.}, the probability of deciding $\mathcal{H}_0$ when $\mathcal{H}_1$ is true) as $N$ grows without bound, under a Neyman-Pearson criterion that the false alarm probability ({\it i.e.}, the probability of deciding $\mathcal{H}_1$ when $\mathcal{H}_0$ is true) is fixed as a constant $0 < \alpha < 1$.  Mathematically, the exponential decay rate is given by
\begin{eqnarray}
\lim_{N \rightarrow \infty} -\frac{1}{N}\log \mbox{Prob}[\delta_N(\underline{Y}) = \mathcal{H}_0|\mathcal{H}_1].
\end{eqnarray}
Likelihood ratio tests (LRTs) achieve the maximal exponential decay rate, and we call this maximum the error exponent. For the detection problem (\ref{eqn:problem}), the error exponent is given by (see, {\it e.g.}, \cite{sung06:it} and references therein)
\begin{eqnarray}
\label{eqn:exponent}
\Gamma = \frac{1}{4\pi} \int_{-\pi}^\pi \left[\log(1 + \phi(\omega)/\sigma^2) - \frac{\phi(\omega)/\sigma^2}{1 + \phi(\omega)/\sigma^2}\right] d\omega,
\end{eqnarray}
for every $0 < \alpha < 1$. Indeed, $\Gamma$ is the limit of a (normalized) Kullback-Leibler distance, $(1/N)D(p_{N, 0}\|p_N)$, as $N$ grows without bound, where $p_{N, 0}$ denotes the $N$-dimensional probability density function (PDF) of $\underline{Y}$ under $\mathcal{H}_0$, and $p_N$ denotes the $N$-dimensional PDF of $\underline{Y}$ under $\mathcal{H}_1$, induced by the signal PSD $\phi(\omega)$.

In order to achieve (\ref{eqn:exponent}), a sequence of LRT detectors (or frequency-domain correlation detectors \cite{zhang10:sp}) need to be built with the exact knowledge of the signal PSD $\phi(\omega)$, $\omega \in [-\pi, \pi]$. Due to practical limitations, however, the knowledge of $\phi(\omega)$ may usually be imprecise. Under such modeling uncertainty, the signal PSD $\phi(\cdot)$ is known only to be within a set $\mathcal{U}_\phi$ of PSD functions. Hence, neither the LRT detectors nor the frequency-domain correlation detectors can be implemented, and it is usually desirable to design robust detectors according to a minimax criterion (see, {\it e.g.}, \cite{kassam85:pieee}). The philosophy of the minimax criterion is as follows. The engineer first chooses a sequence of detectors $\underline{\delta}$, and Nature subsequently responds with a model from $\mathcal{U}_\phi$, which leads to the worst performance for that sequence of detectors. The engineer's task, naturally, is to choose $\underline{\delta}$ such that the resulting worst performance is optimized. As previously discussed, the performance in this paper is the exponential decay rate of the miss detection probability under a fixed false alarm probability. So we can define the minimax robust error exponent as
\begin{eqnarray}
\label{eqn:MR-problem}
\Gamma_{\mathrm{MR}} = \max_{\underline{\delta} \in \Delta_\alpha} \inf_{\phi(\cdot)\in \mathcal{U}_\phi} \lim_{N \rightarrow \infty} -\frac{1}{N}\log\mbox{Prob}[\delta_N(\underline{Y}) = \mathcal{H}_0|\mathcal{H}_1],
\end{eqnarray}
where $\Delta_\alpha$ denotes the set of all the detector sequences that achieve a fixed false alarm probability $0 < \alpha < 1$.

An upper bound to the minimax robust error exponent $\Gamma_{\mathrm{MR}}$ is
\begin{eqnarray}
\label{eqn:upperbound}
\bar{\Gamma}_{\mathrm{MR}} = \inf_{\phi(\cdot) \in \mathcal{U}_\phi} \frac{1}{4\pi} \int_{-\pi}^\pi \left[\log(1 + \phi(\omega)/\sigma^2) - \frac{\phi(\omega)/\sigma^2}{1 + \phi(\omega)/\sigma^2}\right] d\omega,
\end{eqnarray}
assuming that a genie provides the actual $\phi(\cdot)$ to the engineer before choosing the detector sequence.

Generally speaking, genie-aided upper bounds as in (\ref{eqn:upperbound}) are not achievable for minimax robustness problems, unless the problem possesses certain structural properties, say, possessing a saddle point; see, {\it e.g.}, \cite{kassam85:pieee} and references therein for various formulations and approaches to such minimax robustness problems. A minimax robustness problem can be viewed as a game between two players \cite{verdu84:it}, one choosing a filter, which is the sequence of detectors $\underline{\delta}$ here, and the other choosing an operating point which is the model realization $\phi(\cdot)$ here. Typically, the existence of saddle points requires the space of operating points be a convex set; see, {\it e.g.}, \cite{verdu84:it}. Robust decision problems have been studied extensively under various criteria like Bayes risk, error probabilities, generalized signal-to-noise ratio, {\it etc.}; see, {\it e.g.}, \cite{huber65:ams}-\cite{barton90:it}. The minimax robustness problem (\ref{eqn:MR-problem}) regarding the error exponent of detecting a stationary Gaussian process with PSD uncertainty in white Gaussian noise has been studied (among other more general problems) in \cite{kazakos82:it} and \cite{geraniotis85:it}, where sufficient conditions are presented under which exponential decay rates of false alarm and miss probabilities are guaranteed. For hypothesis testing problems in which candidate hypotheses are characterized by moment classes, the asymptotic minimax robustness was investigated in \cite{pandit04:isit}.

In this paper, we establish that, under a dominance condition among the PSDs in $\mathcal{U}_\phi$, the minimax robust error exponent problem (\ref{eqn:MR-problem}) possesses a saddle point solution, and the minimax robust error exponent is achievable by the LRT detectors with respect to the so-called dominated PSD. The dominance condition appears to be a novel property, and imposes no requirement on the convexity of $\mathcal{U}_\phi$.

The main result of our paper is the following.
%\begin{thm}
%\label{thm:MR-value}
%For an arbitrary uncertainty set $\mathcal{U}_\phi$, the minimax robust error exponent is
%\begin{eqnarray}
%\label{eqn:MR-value}
%\Gamma_{\mathrm{MR}} = \inf_{\phi(\cdot) \in \mathcal{U}_\phi} \frac{1}{4\pi} \int_{-\pi}^\pi \left[\log(1 + \phi(\omega)/\sigma^2) - \frac{\phi(\omega)/\sigma^2}{1 + \phi(\omega)/\sigma^2}\right] d\omega.
%\end{eqnarray}
%\end{thm}
\begin{thm}
\label{thm:MR-LLR}
For an arbitrary uncertainty set $\mathcal{U}_\phi$, if there exists a PSD $\phi^\ast(\cdot) \in \mathcal{U}_\phi$, such that the following dominance condition
\begin{eqnarray}
\label{eqn:PSD-dominance-MR-LLR}
\frac{1}{2\pi} \int_{-\pi}^\pi \log\left(1 + \frac{\phi^\ast(\omega)[\phi(\omega) - \phi^\ast(\omega)]}{[\sigma^2 + \phi^\ast(\omega)]^2}\right)d\omega \geq 0
\end{eqnarray}
holds for every $\phi(\cdot) \in \mathcal{U}_\phi$, then the sequence of LRT detectors with respect to $\phi^\ast(\cdot)$ achieves the minimax robust error exponent
\begin{eqnarray}
\label{eqn:MR-value}
\Gamma_{\mathrm{MR}} = \frac{1}{4\pi} \int_{-\pi}^\pi \left[\log(1 + \phi^\ast(\omega)/\sigma^2) - \frac{\phi^\ast(\omega)/\sigma^2}{1 + \phi^\ast(\omega)/\sigma^2}\right] d\omega,
\end{eqnarray}
for every false alarm probability $0 < \alpha < 1$. If $\phi^\ast(\cdot)$ exists, then it is unique.
\end{thm}

Theorem \ref{thm:MR-LLR} is surprising, since the dominance condition (\ref{eqn:PSD-dominance-MR-LLR}) does not require $\mathcal{U}_\phi$ be convex, or generated by $2$-alternating capacities (see, {\it e.g.}, \cite{poor82:ap}), or described by moment classes (see, {\it e.g.}, \cite{pandit04:isit}). Also, when $\phi^\ast(\cdot)$ exists, Theorem \ref{thm:MR-LLR} not only suggests the existence of, but also explicitly gives, the sequence of detectors that achieve $\Gamma_{\mathrm{MR}}$. Furthermore, due to the concavity of logarithmic functions, it follows that, the sequence of LRT detectors with respect to $\phi^\ast(\cdot)$ also solves the minimax robustness problem (\ref{eqn:MR-problem}) when the uncertainty set $\mathcal{U}_\phi$ is enlarged to its convex hull, yielding the same minimax robust error exponent $\Gamma_{\mathrm{MR}}$.

{\it Exemplifications of the dominance condition:}
\begin{itemize}
\item Define the lower envelope function of $\mathcal{U}_\phi$ by $E_l(\omega) = \inf\{\phi(\omega): \forall \phi(\omega) \in \mathcal{U}_\phi\}$, $\omega \in [-\pi, \pi]$. If $E_l(\cdot) \in \mathcal{U}_\phi$, then it is $\phi^\ast(\cdot)$.
\item If $\phi(\omega) = \rho \sigma^2 \in \mathcal{U}_\phi$, and every element of $\mathcal{U}_\phi$ satisfies
\begin{eqnarray}
\frac{1}{2\pi}\int_{-\pi}^\pi\log\left[\frac{\phi(\omega)}{\sigma^2} + \frac{1 + 2\rho}{\rho}\right]d\omega \geq \log\frac{(1 + \rho)^2}{\rho},
\end{eqnarray}
then $\phi^\ast(\omega) = \rho \sigma^2$.
\item If $\sigma^2$ is substantially larger than all the elements of $\mathcal{U}_\phi$ uniformly, {\it i.e.}, very low signal-to-noise ratio, then (\ref{eqn:PSD-dominance-MR-LLR}) is approximated as
\begin{eqnarray}
\frac{1}{2\pi\sigma^4} \int_{-\pi}^\pi \phi^\ast(\omega)\left[\phi(\omega) - \phi^\ast(\omega)\right]d\omega \geq 0,
\end{eqnarray}
which leads to
\begin{eqnarray}
\int_{-\pi}^\pi \left[\phi^\ast(\omega)\right]^2 d\omega \leq \int_{-\pi}^\pi \phi^\ast(\omega) \phi(\omega) d\omega,
\end{eqnarray}
as the criterion for $\phi^\ast(\cdot)$.
\end{itemize}

We devote the remaining parts of this paper to the proof of Theorem \ref{thm:MR-LLR}. In this section, we outline the key ideas in the proof as follows. We start with an arbitrary finite number, $K$, of PSD functions, one of them being a $\phi^\ast(\cdot)$ satisfying (\ref{eqn:PSD-dominance-MR-LLR}), and the other $K - 1$ arbitrarily sampled from $\mathcal{U}_\phi$, denoted $\{\phi_1(\cdot), \phi_2(\cdot), \ldots, \phi_K(\cdot)\}$ where we let $\phi_1(\cdot) = \phi^\ast(\cdot)$. Then, instead of these $K$ isolated PSD functions, we ``convexify'' the problem and consider, for each $N$, the set of mixture probability distributions over the $K$ $N$-dimensional Gaussian distributions induced by $\{\sigma^2 + \phi_1(\cdot), \sigma^2 + \phi_2(\cdot), \ldots, \sigma^2 + \phi_K(\cdot)\}$. Since this set of mixtures is a convex set, we exploit results in the minimax robustness theory (see, {\it e.g.}, \cite{verdu84:it}) to establish that, under the dominance condition (\ref{eqn:PSD-dominance-MR-LLR}), the following Kullback-Leibler distance
\begin{eqnarray}
\label{eqn:mixture-KL-distance}
\min_{\underline{r} \in \mathcal{P}} \frac{1}{N} D\left(p_{N, 0}\left\|\sum_{k=1}^K r_k p_{N, k}\right.\right)
\end{eqnarray}
is achievable as a lower bound to the error exponent as $N$ grows sufficiently large. Here, $p_{N, 0}$ denotes the $N$-dimensional Gaussian noise distribution under $\mathcal{H}_0$, $p_{N, k}$ denotes the $N$-dimensional Gaussian distribution induced by $\sigma^2 + \phi_k(\cdot)$, $\underline{r} \in [0, 1]^K$ denotes the $K$-dimensional mixture vector satisfying $\sum_{k=1}^K r_k = 1$, and $\mathcal{P}$ is the set of all mixture vectors. Now, as $N$ grows without bound, we show that, under the condition (\ref{eqn:PSD-dominance-MR-LLR}), the value of (\ref{eqn:mixture-KL-distance}) converges to
\begin{eqnarray}
\label{eqn:KL-distance-K-limit}
\lim_{N \rightarrow \infty} \frac{1}{N} D\left(p_{N, 0}\left\| p_{N, 1}\right.\right),
\end{eqnarray}
which we see is the minimax robust error exponent, $\Gamma_{\mathrm{MR}}$ by noting that $\phi_1(\cdot) = \phi^\ast(\cdot)$. Since we have established that $\Gamma_{\mathrm{MR}}$ is achievable over the set of mixture distributions, it is also achievable over the smaller set of $K$ $N$-dimensional Gaussian distributions induced by the $K$ PSD functions $\{\sigma^2 + \phi_1(\cdot), \sigma^2 + \phi_2(\cdot), \ldots, \sigma^2 + \phi_K(\cdot)\}$. From the above procedure, we establish that for every $K$-point set $\{\phi_1(\cdot), \phi_2(\cdot), \ldots, \phi_K(\cdot)\}$ where $\phi_1(\cdot) = \phi^\ast(\cdot)$, there exists a sequence of detectors that achieves $\Gamma_{\mathrm{MR}}$ over that $K$-point set.

Although the Kullback-Leibler distance with respect to the mixture distribution, (\ref{eqn:mixture-KL-distance}), converges to (\ref{eqn:KL-distance-K-limit}), this by no means implies that the minimax robust detector sequence asymptotically converges to the LRT for Gaussian distributions. Generally speaking, the minimax robust detector sequence may be the LRT for a sequence of mixture Gaussian distributions. In order for the minimax robust detector sequence to be the LRT for Gaussian distributions, it is necessary for the solution of the minimization problem (\ref{eqn:mixture-KL-distance}) to be a ``singleton'', {\it i.e.}, all but one component of $\underline{r}$ are zeros. Applying the Karush-Kuhn-Tucker (KKT) conditions, we show that the dominance condition (\ref{eqn:PSD-dominance-MR-LLR}) also warrants that the minimization problem (\ref{eqn:mixture-KL-distance}) can be solved by the sequence of LRT detectors with respect to $\phi^\ast(\cdot)$, hence concluding the proof of Theorem \ref{thm:MR-LLR}.

%%%%%%%%%%%%%%%%%%%%%%%%%%%%%%%%%%%%%%%%%%%%%%%%%%%%%%%%%%%%%%%%%%%%%%%%%%%%%%%%%%%%%%%%%%%%%%%%%%%%%%%%%%%%
\section{PSD Uncertainty Set that Possesses a Dominance Structure}
\label{sec:dominance}

For a given sample space $\Omega$ and its associated $\sigma$-algebra $\mathcal{F}$, we start with three arbitrary probability measures $P_0$, $P_1$ and $P_2$, in which both $P_1$ and $P_2$ are absolutely continuous with respect to $P_0$. We define a dominance relation as follows.
\begin{defn}
\label{defn:dominance-prob}
If it holds that
\begin{eqnarray}
\int_\Omega \frac{dP_2/dP_0}{dP_1/dP_0} dP_0 \leq 1,
\end{eqnarray}
then $P_1$ is dominated by $P_2$ with respect to $P_0$, a condition denoted by $P_1 \stackrel{P_0}{\prec} P_2$. We call $P_0$ the reference probability measure.
\end{defn}

The following lemma immediately follows from Definition \ref{defn:dominance-prob}.
\begin{lem}
\label{lem:never-both}
Unless $dP_1/dP_0$ and $dP_2/dP_0$ are $P_0$-almost surely equal, the two dominance relationships $P_1 \stackrel{P_0}{\prec} P_2$ and $P_2 \stackrel{P_0}{\prec} P_1$ cannot simultaneously hold.
\end{lem}
{\it Proof:} We prove Lemma \ref{lem:never-both} by contradiction. Assume $P_1 \stackrel{P_0}{\prec} P_2$ and $P_2 \stackrel{P_0}{\prec} P_1$ hold simultaneously; that is,
\begin{eqnarray*}
\int_\Omega \frac{dP_2/dP_0}{dP_1/dP_0} dP_0 \leq 1,\\
\int_\Omega \frac{dP_1/dP_0}{dP_2/dP_0} dP_0 \leq 1.
\end{eqnarray*}
Summing these two inequalities leads to
\begin{eqnarray}
\label{eqn:lemma-never-both-proof-1}
\int_\Omega \left(\frac{dP_1/dP_0}{dP_2/dP_0} + \frac{dP_2/dP_0}{dP_1/dP_0}\right) dP_0 \leq 2.
\end{eqnarray}
In (\ref{eqn:lemma-never-both-proof-1}), however, the left hand side is lower bounded by
\begin{eqnarray}
\int_\Omega \left(\frac{dP_1/dP_0}{dP_2/dP_0} + \frac{dP_2/dP_0}{dP_1/dP_0}\right) dP_0 &=& \int_\Omega \left[\left(\sqrt{\frac{dP_1/dP_0}{dP_2/dP_0}} - \sqrt{\frac{dP_2/dP_0}{dP_1/dP_0}}\right)^2 + 2\right] dP_0\nonumber\\
&\geq& 2 \int_\Omega dP_0 = 2.
\end{eqnarray}
Hence the only possible case is where $dP_1/dP_0 = dP_2/dP_0$ except on a subset of $\Omega$ whose $P_0$-measure is zero. But this case has already been excluded in the condition. So we arrive at a contradiction and Lemma \ref{lem:never-both} is established. $\Box$

Now for a set of probability measures, we can define its dominance property if it contains an element probability measure that is dominated by all the others in the set.
\begin{defn}
\label{defn:dominance-prob-set}
Consider the sample space $\Omega$, its associated $\sigma$-algebra $\mathcal{F}$, a reference probability measure $P_0$, and a set of probability measures $\mathcal{P}$. A probability measure $P^\ast \in \mathcal{P}$ is dominated by $\mathcal{P}$ with respect to $P_0$ if for every $P \in \mathcal{P}$, $P^\ast \stackrel{P_0}{\prec} P$. We denote the dominance relationship by $P^\ast \stackrel{P_0}{\prec} \mathcal{P}$.
\end{defn}

In light of Lemma \ref{lem:never-both}, the following lemma is immediate.
\begin{lem}
\label{lem:uniqueness}
A probability measure $P^\ast \in \mathcal{P}$ that is dominated by $\mathcal{P}$ with respect to $P_0$, if it exists, is unique.
\end{lem}

We also note that in general a set of probability measures $\mathcal{P}$ may not contain a dominated element. As a simple example, consider a binary sample space $\Omega = \{0, 1\}$, over which $P_0$ is given by the probability mass function (PMF) $P_0(0) = P_0(1) = 0.5$. For the set $\mathcal{P}$ of two PMF's:
\begin{eqnarray*}
P_1(0) = 0.9,\;P_1(1) = 0.1; \quad P_2(0) = 0.1,\;P_2(1) = 0.9,
\end{eqnarray*}
it is easily verified that neither $P_1$ nor $P_0$ is dominated.

For $\Omega = \mathbb{R}^N$, consider $N$-dimensional zero-mean Gaussian distributions. Fix the reference probability measure $P_0 \sim \mathcal{N}(0, \sigma^2 \mathbf{I}_{N \times N})$. Consider $P_1 \sim \mathcal{N}(0, \mathbf{\Sigma}_1)$ and $P_2 \sim \mathcal{N}(0, \mathbf{\Sigma}_2)$. From Definition \ref{defn:dominance-prob}, let us examine the conditions for $\mathcal{N}(0, \mathbf{\Sigma}_1) \stackrel{\mathcal{N}(0, \sigma^2 \mathbf{I})}{\prec} \mathcal{N}(0, \mathbf{\Sigma}_2)$. We have,
\begin{eqnarray}
&&\mathcal{N}(0, \mathbf{\Sigma}_1) \stackrel{\mathcal{N}(0, \sigma^2 \mathbf{I})}{\prec} \mathcal{N}(0, \mathbf{\Sigma}_2) \Leftrightarrow\nonumber\\
&&\int_{\mathbb{R}^N} \frac{\frac{1}{(2\pi)^{N/2} |\mathbf{\Sigma}_2|^{1/2}}\exp\left[-\frac{1}{2}\underline{x}^T \mathbf{\Sigma}_2^{-1} \underline{x}\right]}{\frac{1}{(2\pi)^{N/2} |\mathbf{\Sigma}_1|^{1/2}}\exp\left[-\frac{1}{2}\underline{x}^T \mathbf{\Sigma}_1^{-1} \underline{x}\right]} \frac{1}{(2\pi\sigma^2)^{N/2}}\exp\left[-\frac{1}{2\sigma^2}\underline{x}^T \underline{x}\right] d\underline{x} \leq 1\nonumber\\
\Rightarrow && \int_{\mathbb{R}^N} \frac{1}{(2\pi\sigma^2)^{N/2}} \frac{|\mathbf{\Sigma}_1|^{1/2}}{|\mathbf{\Sigma}_2|^{1/2}} \exp\left[-\frac{1}{2}\underline{x}^T \left(\mathbf{\Sigma}_2^{-1} - \mathbf{\Sigma}_1^{-1} + \frac{1}{\sigma^2}\mathbf{I}\right)\underline{x}\right]d\underline{x} \leq 1\nonumber\\
\Rightarrow && \left[\frac{|\mathbf{\Sigma}_1|}{|\mathbf{\Sigma}_2|\cdot|\mathbf{I} + \sigma^2(\mathbf{\Sigma}_2^{-1} - \mathbf{\Sigma}_1^{-1})|}\right]^{1/2} \leq 1\nonumber\\
\Rightarrow && |\mathbf{\Sigma}_2^{-1} \mathbf{\Sigma}_1| \leq |\mathbf{I} + \sigma^2\left(\mathbf{\Sigma}_2^{-1} - \mathbf{\Sigma}_1^{-1}\right)|.
\end{eqnarray}
In the above steps, it is implicitly required that the matrix $\mathbf{I} + \sigma^2 \left(\mathbf{\Sigma}_2^{-1} - \mathbf{\Sigma}_1^{-1}\right)$ is positive definite, in order to ensure the convergence of the integral. So we have the following two conditions for $\mathcal{N}(0, \mathbf{\Sigma}_1) \stackrel{\mathcal{N}(0, \sigma^2 \mathbf{I})}{\prec} \mathcal{N}(0, \mathbf{\Sigma}_2)$:
\begin{eqnarray}
\label{eqn:Ndim-cond-pd}
\mathbf{I} + \sigma^2 \left(\mathbf{\Sigma}_2^{-1} - \mathbf{\Sigma}_1^{-1}\right) \;\mbox{is positive definite};\\
\label{eqn:Ndim-cond-det}
\mbox{and}\;\;|\mathbf{\Sigma}_2^{-1}\mathbf{\Sigma}_1| \leq |\mathbf{I} + \sigma^2 \left(\mathbf{\Sigma}_2^{-1} - \mathbf{\Sigma}_1^{-1}\right)|.
\end{eqnarray}

Now as $N$ grows without bound, consider a WSS zero-mean Gaussian process $\{Y_n\}$ with two possible PSDs $\phi_1(\omega)$ and $\phi_2(\omega)$, $\omega \in [-\pi, \pi]$. Denote the probability measures of a length-$N$ segment of $\{Y_n\}$ under $\phi_1(\cdot)$ and $\phi_2(\cdot)$ by $P_{N, 1}$ and $P_{N, 2}$, respectively. Applying the asymptotic properties of Toeplitz matrices (see, {\it e.g.}, \cite[Thm. 5.4]{gray06:book}) to the conditions (\ref{eqn:Ndim-cond-pd})-(\ref{eqn:Ndim-cond-det}), we find that if
\begin{eqnarray}
\label{eqn:psd-dom-cond-1}
\frac{1}{2\pi} \int_{-\pi}^\pi \log\frac{\phi_1(\omega)}{\phi_2(\omega)} d\omega \leq \frac{1}{2\pi}\int_{-\pi}^\pi \log\left(1 + \frac{\sigma^2}{\phi_2(\omega)} - \frac{\sigma^2}{\phi_1(\omega)}\right)d\omega
\end{eqnarray}
holds, and $1 + \sigma^2/\phi_2(\omega) - \sigma^2/\phi_1(\omega)$ is bounded away from zero for all $\omega \in [-\pi, \pi]$, then $P_{N, 1}$ is dominated by $P_{N, 2}$ with respect to $\mathcal{N}(0, \sigma^2\mathbf{I}_{N \times N})$ for all sufficiently large $N$. The condition (\ref{eqn:psd-dom-cond-1}) can further be rewritten as
\begin{eqnarray}
\frac{1}{2\pi}\int_{-\pi}^\pi \log\left(\frac{\sigma^2(\phi_1(\omega) - \phi_2(\omega))}{\phi_1^2(\omega)} + \frac{\phi_2(\omega)}{\phi_1(\omega)}\right)d\omega \geq 0.
\end{eqnarray}

\begin{defn}
\label{defn:dominance-psd}
Consider a set of PSD functions $\mathcal{S}$. A PSD $\phi^\ast(\cdot) \in \mathcal{S}$ is $\sigma^2$-dominated by $\mathcal{S}$ if for every $\phi(\cdot) \in \mathcal{S}$, $1 + \sigma^2/\phi(\omega) - \sigma^2/\phi^\ast(\omega)$ is bounded away from zero for all $\omega \in [-\pi, \pi]$, and
\begin{eqnarray}
\frac{1}{2\pi}\int_{-\pi}^\pi \log\left(\frac{\sigma^2(\phi^\ast(\omega) - \phi(\omega))}{[\phi^\ast(\omega)]^2} + \frac{\phi(\omega)}{\phi^\ast(\omega)}\right)d\omega \geq 0.
\end{eqnarray}
We denote the $\sigma^2$-dominance relationship by $\phi^\ast(\cdot) \stackrel{\sigma^2}{\prec} \mathcal{S}$.
\end{defn}

For the purposes of this paper, we further focus on the $\sigma^2$-translation of the PSD set $\mathcal{S}$, obtained by adding a noise floor of $\sigma^2$ to each element PSD of $\mathcal{S}$; that is,
\begin{eqnarray}
\mathcal{S}_{[\sigma^2]} = \left\{\sigma^2 + \phi(\cdot): \phi(\cdot) \in \mathcal{S}\right\}.
\end{eqnarray}

According to Definition \ref{defn:dominance-psd}, $\sigma^2 + \phi^\ast(\cdot)\in \mathcal{S}_{[\sigma^2]}$ is dominated by $\mathcal{S}_{[\sigma^2]}$ if for every $\phi(\cdot) \in \mathcal{S}$,
\begin{eqnarray}
\frac{1}{2\pi} \int_{-\pi}^\pi \log\left(1 + \frac{\phi^\ast(\omega)\left[\phi(\omega) - \phi^\ast(\omega)\right]}{\left[\sigma^2 + \phi^\ast(\omega)\right]^2}\right)d\omega \geq 0.
\end{eqnarray}
This is the same as the dominance condition (\ref{eqn:PSD-dominance-MR-LLR}) in Theorem \ref{thm:MR-LLR}.

%%%%%%%%%%%%%%%%%%%%%%%%%%%%%%%%%%%%%%%%%%%%%%%%%%%%%%%%%%%%%%%%%%%%%%%%%%%%%%%%%%%%%%%%%%%%%%%%%%%%%%%%%%%%
\section{Asymptotic Behavior of Detector Sequences Based on Gaussian Mixtures}
\label{sec:detector-mixture-behavior}

Consider an arbitrary finite number, $K$, of possible PSD functions,
\begin{eqnarray}
\mathcal{S} = \{\phi_1(\cdot), \phi_2(\cdot), \ldots, \phi_K(\cdot)\}.
\end{eqnarray}
In $\mathcal{S}$, we always keep the dominated PSD $\phi^\ast(\cdot)$ and index it as $\phi_1(\cdot)$;\footnote{We shall use these two notations interchangeably in the sequel.} the other $(K - 1)$ PSD functions are arbitrarily sampled from $\mathcal{U}_\phi$. For convenience, denote the $\left[\sigma^2 + \phi_k(\cdot)\right]$-induced covariance matrix of the $N$-dimensional Gaussian distribution by $\sigma^2 \mathbf{I}_{N \times N} + \mathbf{\Sigma}_{N, k}$, and its PDF by $p_{N, k}$:
\begin{eqnarray}
p_{N, k}(\underline{y}) = \frac{1}{(2\pi)^{N/2} |\sigma^2 \mathbf{I} + \mathbf{\Sigma}_{N, k}|^{1/2}} \exp\left[-(1/2)\underline{y}^T \left(\sigma^2 \mathbf{I} + \mathbf{\Sigma}_{N, k}\right)^{-1}\underline{y}\right].
\end{eqnarray}
Also denote by $p_{N, 0}$ the PDF of the $N$-dimensional Gaussian distribution under $\mathcal{H}_0$:
\begin{eqnarray}
p_{N, 0}(\underline{y}) = \frac{1}{(2\pi\sigma^2)^{N/2}} \exp\left[-\underline{y}^T \underline{y}/(2\sigma^2)\right].
\end{eqnarray}

We consider detector sequences whose decision statistics take the following form:
\begin{eqnarray}
\label{eqn:g-decision-statistics}
g_N(\underline{y}; \underline{q}) &=& \frac{1}{N}\log\sum_{k = 1}^K q_k \frac{p_{N, k}(\underline{y})}{p_{N, 0}(\underline{y})}\nonumber\\
&=& \frac{1}{N}\log\sum_{k = 1}^K \frac{q_k}{|\mathbf{I} + \mathbf{\Sigma}_{N, k}/\sigma^2|^{1/2}} \exp\left[\frac{1}{2\sigma^2} \underline{y}^T (\sigma^2 \mathbf{I} + \mathbf{\Sigma}_{N, k})^{-1}\mathbf{\Sigma}_{N, k}\underline{y}\right],
\end{eqnarray}
where the vector $\underline{q}$ satisfies the normalization condition $\sum_{k = 1}^K q_k = 1$, $q_1 > 0$, and $q_k \geq 0$, $\forall k \neq 1$. Note that we restrict the component $q_1$ corresponding to $\phi^\ast(\cdot)$ to be strictly positive. Indexed by $N$, the considered sequence of detectors are deterministic threshold tests,
\begin{eqnarray}
\mbox{if}\; g_N(\underline{y}; \underline{q}) \leq \tau_N,\; \delta_N(\underline{y}) = \mathcal{H}_0;\;\mbox{otherwise,}\; \delta_N(\underline{y}) = \mathcal{H}_1
\end{eqnarray}
with thresholds $\{\tau_N\}$. For each $N$, $\tau_N$ is determined through the constraint that the false alarm probability is fixed as $\alpha$, {\it i.e.},
\begin{eqnarray}
\int_{g_N(\underline{y}; \underline{q}) > \tau_N} p_{N, 0}(\underline{y}) d\underline{y} = \alpha.
\end{eqnarray}

In this section, we investigate the asymptotic behavior of $g_N(\underline{Y}; \underline{q})$ when $\underline{Y}$ follows $\mathcal{H}_0$, as $N$ grows large.

For each $k = 1, 2, \ldots, K$, we can define a quantity
\begin{eqnarray}
\psi_k = \frac{1}{4\pi} \int_{-\pi}^\pi \left[\log\left(1 +\phi_k(\omega)/\sigma^2\right) - \frac{\phi_k(\omega)/\sigma^2}{1 + \phi_k(\omega)/\sigma^2}\right]d\omega.
\end{eqnarray}
The first step in our investigation, Lemma \ref{lem:phi-ast-minimizer}, indicates that $\phi^\ast(\cdot)$ attains $\min_{k} \psi_k$.

\begin{lem}
\label{lem:phi-ast-minimizer}
\begin{eqnarray}
\min_k \psi_k = \frac{1}{4\pi} \int_{-\pi}^\pi \left[\log\left(1 +\phi^\ast(\omega)/\sigma^2\right) - \frac{\phi^\ast(\omega)/\sigma^2}{1 + \phi^\ast(\omega)/\sigma^2}\right]d\omega.
\end{eqnarray}
\end{lem}
{\it Proof:} As we have noted regarding the error exponent $\Gamma$ in the introduction, $\psi_k$ is the limit of
\begin{eqnarray*}
\frac{1}{N} D\left(p_{N, 0}\|p_{N, k}\right).
\end{eqnarray*}
So in order to prove the result, we consider the following related ``convexified'' problem
\begin{eqnarray}
\label{eqn:phi-ast-minimizer-convexified}
\min_{\underline{r}} \frac{1}{N} D\left(p_{N, 0}\left\|\sum_{k = 1}^K r_k p_{N, k} \right.\right),
\end{eqnarray}
where $\underline{r} \in [0, 1]^K$ satisfies $\sum_{k = 1}^K r_k = 1$. If we prove that for every sufficiently large $N$, the $N$-dimensional probability distribution induced by $\phi^\ast(\cdot)$ solves (\ref{eqn:phi-ast-minimizer-convexified}), then it also solves the original problem of minimizing $\psi_k$.

Since the Kullback-Leibler distance is convex with respect to its operand distributions, the KKT conditions \cite{luenberger69:book} provide necessary and sufficient conditions for optimality, as
\begin{eqnarray}
\label{eqn:KL-minimizer-KKT-1}
\frac{1}{N}\mathbf{E}_{p_{N, 0}}\left[\frac{p_{N, k}}{\sum_{k = 1}^K r_k p_{N, k}}\right] + \mu_k - \lambda = 0, \quad \forall k;\\
\mu_k \geq 0, \quad \forall k;\\
\mu_k r_k = 0, \quad \forall k.
\end{eqnarray}
So, if $\underline{r} = [1, 0, 0, \ldots, 0]$ ({\it i.e.}, $\phi^\ast(\cdot)$) is the minimizer of $(1/N) D\left(p_{N, 0}\left\|\sum_{k = 1}^K r_k p_{N, k} \right.\right)$, we have $\mu_1 = 0$, which when substituted into (\ref{eqn:KL-minimizer-KKT-1}) leads to
\begin{eqnarray}
\frac{1}{N} \mathbf{E}_{p_{N, 0}}\left[\frac{p_{N, 1}}{p_{N, 1}}\right] + 0 - \lambda = 0 \Rightarrow \lambda = \frac{1}{N}.
\end{eqnarray}
Substituting $\lambda = 1/N$ into (\ref{eqn:KL-minimizer-KKT-1}) for $k \neq 1$ leads to
\begin{eqnarray}
\mathbf{E}_{p_{N, 0}}\left[\frac{p_{N, k}}{p_{N, 1}}\right] - 1 = -N\mu_k \leq 0, \quad\forall k \neq 1.
\end{eqnarray}
As $N$ grows large, this results in the dominance condition (\ref{eqn:PSD-dominance-MR-LLR}), according to the development in Section \ref{sec:dominance}. So Lemma \ref{lem:phi-ast-minimizer} is established. $\Box$

The following two lemmas then characterize the asymptotic behavior of $g_N(\underline{Y}; \underline{q})$ under $\mathcal{H}_0$.
\begin{lem}
\label{lem:threshold-in-prob}
Under the dominance condition (\ref{eqn:PSD-dominance-MR-LLR}), for every $\underline{q}$ with $q_1 \neq 0$, the threshold sequence $\{\tau_N\}$ converges to $- \frac{1}{4\pi} \int_{-\pi}^\pi \left[\log\left(1 +\phi^\ast(\omega)/\sigma^2\right) - \frac{\phi^\ast(\omega)/\sigma^2}{1 + \phi^\ast(\omega)/\sigma^2}\right]d\omega$ as $N \rightarrow \infty$, for every fixed false alarm probability $0 < \alpha < 1$.
\end{lem}
\begin{lem}
\label{lem:threshold-expectation}
Under the dominance condition (\ref{eqn:PSD-dominance-MR-LLR}), for every $\underline{q}$ with $q_1 \neq 0$, the decision statistics satisfy
\begin{eqnarray}
\label{eqn:threshold-limit-expectation}
\lim_{N \rightarrow\infty} \mathbf{E}_{p_{N, 0}}\left[g_N(\underline{Y}; \underline{q})\right] = - \frac{1}{4\pi} \int_{-\pi}^\pi \left[\log\left(1 +\phi^\ast(\omega)/\sigma^2\right) - \frac{\phi^\ast(\omega)/\sigma^2}{1 + \phi^\ast(\omega)/\sigma^2}\right]d\omega.
\end{eqnarray}
\end{lem}
{\it Proof of Lemma \ref{lem:threshold-in-prob}:} In light of Lemma \ref{lem:phi-ast-minimizer}, the result would straightforwardly follow if we prove that under $\mathcal{H}_0$, the sequence of decision statistics, $\{g_N(\underline{Y}; \underline{q})\}$, converges to $-\min_{k} \psi_k$ in probability. We do so using a sandwich type of proof technique. On one hand, $g_N(\underline{Y}; \underline{q})$ is lower bounded as
\begin{eqnarray}
\label{eqn:threshold-in-prob-proof-1}
g_N(\underline{Y}; \underline{q}) &=& \frac{1}{N} \log \sum_{k = 1}^K q_k \frac{p_{N, k}(\underline{Y})}{p_{N, 0}(\underline{Y})}\nonumber\\
&\geq& \frac{1}{N} \log q_k \frac{p_{N, k}(\underline{Y})}{p_{N, 0}(\underline{Y})}\nonumber\\
&=& \frac{1}{N}\log \frac{1}{|\mathbf{I} + \mathbf{\Sigma}_{N, k}/\sigma^2|^{1/2}} \exp\left[\frac{1}{2\sigma^2} \underline{Y}^T (\sigma^2 \mathbf{I} + \mathbf{\Sigma}_{N, k})^{-1}\mathbf{\Sigma}_{N, k}\underline{Y}\right] + \frac{\log q_k}{N},
\end{eqnarray}
for every $k$. On noting that under $\mathcal{H}_0$ $\underline{Y}$ is an $N$-dimensional Gaussian random vector with covariance matrix $\sigma^2 \mathbf{I}$, (\ref{eqn:threshold-in-prob-proof-1}) can further be rewritten as
\begin{eqnarray}
\label{eqn:threshold-in-prob-proof-2}
g_N(\underline{Y}; \underline{q}) \geq \frac{1}{N} \left[\sum_{n = 0}^{N-1} \lambda_{k, n} W_n^2 - \sum_{n = 0}^{N-1} \log\mu_{k, n}\right] + \frac{\log q_k}{N},
\end{eqnarray}
where $\lambda_{k, n}$ denotes the $n$-th eigenvalue of $(1/2) (\sigma^2 \mathbf{I} + \mathbf{\Sigma}_{N, k})^{-1}\mathbf{\Sigma}_{N, k}$, $\mu_{k, n}$ denotes the $n$-th eigenvalue of $\left(\mathbf{I} + \mathbf{\Sigma}_{N, k}/\sigma^2\right)^{1/2}$, and $\{W_n\}_{n=0}^{N-1}$ are i.i.d. zero-mean unit-variance Gaussian random variables. Then from Chebyshev's inequality and the asymptotic properties of Hermitian Toeplitz matrices (see, {\it e.g.}, \cite{gray06:book}), $\lim_{N \rightarrow \infty} g_N(\underline{Y}; \underline{q})$ is lower bounded by
\begin{eqnarray*}
- \frac{1}{4\pi} \int_{-\pi}^\pi \left[\log\left(1 +\phi_k(\omega)/\sigma^2\right) - \frac{\phi_k(\omega)/\sigma^2}{1 + \phi_k(\omega)/\sigma^2}\right]d\omega = -\psi_k
\end{eqnarray*}
in probability, for every $k$ such that $q_k \neq 0$. So the tightest lower bound yields $\lim_{N \rightarrow \infty} g_N(\underline{Y}; \underline{q}) \geq -\min_k \psi_k$ in probability.

On the other hand, we wish to prove that for any small $\epsilon > 0$, as $N$ grows without bound, $g_N(\underline{Y}; \underline{q}) \leq -\min_k \psi_k + \epsilon$ with vanishingly small probability. For this, it suffices to prove that for every $k > 1$ with $q_k \neq 0$, $p_{N, k}(\underline{Y})$ is exponentially smaller than $p_{N, 1}(\underline{Y})$ (which is induced by $\sigma^2 + \phi^\ast(\cdot)$) with high probability. This also follows from similar steps as those used in establishing the lower bound above. Consequently, Lemma \ref{lem:threshold-in-prob} is established. $\Box$

{\it Proof of Lemma \ref{lem:threshold-expectation}:} Similar to Lemma \ref{lem:threshold-in-prob}, the proof is also based on a sandwich type of technique. The lower bound of $\mathbf{E}_{p_{N, 0}}[g_N(\underline{Y}; \underline{q})]$ follows essentially the same line as in establishing the lower bound in Lemma \ref{lem:threshold-in-prob}, and we have
\begin{eqnarray}
\lim_{N \rightarrow \infty} \mathbf{E}_{p_{N, 0}}[g_N(\underline{Y}; \underline{q})] \geq -\frac{1}{4\pi} \int_{-\pi}^\pi \left[\log\left(1 +\phi^\ast(\omega)/\sigma^2\right) - \frac{\phi^\ast(\omega)/\sigma^2}{1 + \phi^\ast(\omega)/\sigma^2}\right]d\omega.
\end{eqnarray}
To establish an upper bound, we note that
\begin{eqnarray}
&&\mathbf{E}_{p_{N, 0}}\left[g_N(\underline{Y}; \underline{q})\right] \nonumber\\
&=& \frac{1}{N} \mathbf{E}_{p_{N, 0}} \log \sum_{k = 1}^K q_k \frac{p_{N, k}(\underline{Y})}{p_{N, 0}(\underline{Y})}\nonumber\\
\label{eqn:g-UB}
&=& \frac{1}{N}\mathbf{E}_{p_{N, 0}}\log \left[q_1 \frac{p_{N, 1}(\underline{Y})}{p_{N, 0}(\underline{Y})}\right] + \frac{1}{N} \mathbf{E}_{p_{N, 0}}\log \left[1 + \sum_{k \neq 1, q_k \neq 0} \frac{q_k}{q_1}\frac{p_{N, k}(\underline{Y})}{p_{N, 1}(\underline{Y})}\right].
\end{eqnarray}
In (\ref{eqn:g-UB}), the first term converges to $-\frac{1}{4\pi} \int_{-\pi}^\pi \left[\log\left(1 +\phi^\ast(\omega)/\sigma^2\right) - \frac{\phi^\ast(\omega)/\sigma^2}{1 + \phi^\ast(\omega)/\sigma^2}\right]d\omega$ following the lower bounding procedure; the second term can be upper bounded as
\begin{eqnarray}
\frac{1}{N} \mathbf{E}_{p_{N, 0}}\log \left[1 + \sum_{k \neq 1, q_k \neq 0} \frac{q_k}{q_1}\frac{p_{N, k}(\underline{Y})}{p_{N, 1}(\underline{Y})}\right] \leq \frac{1}{N} \sum_{k \neq 1, q_k \neq 0} \frac{q_k}{q_1} \mathbf{E}_{p_{N, 0}}\left[\frac{p_{N, k}(\underline{Y})}{p_{N, 1}(\underline{Y})}\right],
\end{eqnarray}
since $\log(1 + x) \leq x$ for all $x > -1$. Now it suffices to prove that for every $k \neq 1$, $\mathbf{E}_{p_{N, 0}}\left[\frac{p_{N, k}(\underline{Y})}{p_{N, 1}(\underline{Y})}\right]$ is bounded. From the development of $\sigma^2$-dominance in Section \ref{sec:dominance}, for all sufficiently large $N$, this condition is implied by the dominance condition (\ref{eqn:PSD-dominance-MR-LLR}). This concludes the proof of Lemma \ref{lem:threshold-expectation}. $\Box$

%%%%%%%%%%%%%%%%%%%%%%%%%%%%%%%%%%%%%%%%%%%%%%%%%%%%%%%%%%%%%%%%%%%%%%%%%%%%%%%%%%%%%%%%%%%%%%%%%%%%%%%%%%%%
\section{Proof of Theorem \ref{thm:MR-LLR}}
\label{sec:proof}

In this section, we use the auxiliary results developed in the previous sections to establish Theorem \ref{thm:MR-LLR}. We shall consider the hypothesis testing problem (\ref{eqn:problem}) over an arbitrary $K$-point set of possible PSD functions $\mathcal{S}$, which contains $\phi^\ast(\cdot)$ as $\phi_1(\cdot)$ and another $(K - 1)$ arbitrarily sampled PSD functions from $\mathcal{U}_\phi$. Once we prove that for any such $\mathcal{S}$, LRT detectors with respect to $\phi^\ast(\cdot)$ achieve the minimax robust error exponent $\Gamma_{\mathrm{MR}}$, for every $0 < \alpha < 1$, then Theorem \ref{thm:MR-LLR} straightforwardly follows through a contradiction argument.

According to Chernoff's bound, the miss probability of the decision procedure using $\{g_N(\underline{y}; \underline{q}), \tau_N\}$ is upper bounded by
\begin{eqnarray}
\mbox{Pr}[\delta_N(\underline{Y}) = 0|\mathcal{H}_1] \leq \exp\left\{-N\cdot \sup_{s \leq 0}\left[\frac{s}{N} \tau_N - \frac{1}{N}\log\mathbf{E}_{p_N}\left[e^{s g_N(\underline{Y}; \underline{q})}\right]\right]\right\},
\end{eqnarray}
where the expectation is with respect to $p_N$, the distribution of $\underline{Y}$ under $\mathcal{H}_1$. So from Lemmas \ref{lem:threshold-in-prob} and \ref{lem:threshold-expectation}, for any $\epsilon > 0$, there exists a sufficiently large $N_\epsilon$, such that for all $N > N_\epsilon$, the Chernoff's bound gives
\begin{eqnarray}
\mbox{Pr}[\delta_N(\underline{Y}) = 0|\mathcal{H}_1] \leq \exp\left\{-N\cdot \sup_{s \leq 0}\left[\frac{s}{N} \left(\mathbf{E}_{p_{N, 0}}\left[g_N(\underline{Y}; \underline{q})\right] + \epsilon\right) - \frac{1}{N}\log\mathbf{E}_{p_N}\left[e^{s g_N(\underline{Y}; \underline{q})}\right]\right]\right\}.
\end{eqnarray}
As $\epsilon \rightarrow 0$, we pose the following minimax problem to optimize the minimax robustness performance of $N$-dimensional decision-making under the Neyman-Pearson criterion:
\begin{eqnarray}
\label{eqn:minimax-v1}
\max_{\underline{q}} \min_{k = 1, 2, \ldots, K} \sup_{s \leq 0}\left[\frac{s}{N}\mathbf{E}_{p_{N, 0}}\left[g_N(\underline{Y}; \underline{q})\right] - \frac{1}{N}\log\mathbf{E}_{p_{N, k}}\left[e^{s g_N(\underline{Y}; \underline{q})}\right]\right].
\end{eqnarray}

To proceed using the minimax robustness theory \cite{verdu84:it}, we augment the sets in problem (\ref{eqn:minimax-v1}). Instead of restricting $q_1$ to be strictly positive, we consider $\underline{q} \in \mathcal{P}$, where
\begin{eqnarray}
\mathcal{P} = \left\{\underline{x} \in [0, 1]^K: \sum_{k = 1}^K x_k = 1\right\}.
\end{eqnarray}
Instead of considering probability distributions induced by the $K$ isolated PSD functions in $\mathcal{S}$, we consider the PDF of $\underline{Y}$ as a convex combination of $\{p_{N, 1}, p_{N, 2}, \ldots, p_{N, K}\}$, as
\begin{eqnarray}
\label{eqn:p_N-mix}
p_N(\underline{y}) \in \left\{\sum_{k = 1}^K r_k p_{N, k}(\underline{y}): \sum_{k = 1}^K r_k = 1; r_k \geq 0, \forall k \right\}.
\end{eqnarray}
Note that in general $p_N$ corresponds to a mixture of $N$-dimensional Gaussian distributions, unless $\underline{r}$ is a ``singleton'', {\it i.e.}, all but one component of $\underline{r}$ are zeros. For convenience, we write (\ref{eqn:p_N-mix}) as $p_N(\underline{y}; \underline{r})$ in order to reflect its dependence on $\underline{r}$. For the detectors sequences with decision statistics $g_N(\underline{y}; \underline{q})$ in the form (\ref{eqn:g-decision-statistics}), when $\underline{q} = \underline{r}$, $g_N(\underline{y}; \underline{r})$ is the log-likelihood ratio test (LLRT) statistic.

Now we consider the augmented minimax problem
\begin{eqnarray}
\label{eqn:minimax-v2}
\max_{\underline{q} \in \mathcal{P}} \min_{\underline{r} \in \mathcal{P}} \sup_{s \leq 0}\left[\frac{s}{N}\mathbf{E}_{p_{N, 0}}\left[g_N(\underline{Y}; \underline{q})\right] - \frac{1}{N}\log\mathbf{E}_{p_N(\cdot; \underline{r})}\left[e^{s g_N(\underline{Y}; \underline{q})}\right]\right].
\end{eqnarray}

Since the sets in (\ref{eqn:minimax-v1}) are subsets of those in (\ref{eqn:minimax-v2}), if we can prove that $p_{N, 1}$ (which is induced by $\phi^\ast(\cdot)$) and its associated LRT solve (\ref{eqn:minimax-v2}), then they also solve (\ref{eqn:minimax-v1}). In the following, we prove the minimax robustness result for (\ref{eqn:minimax-v2}), through following the general approach developed in \cite{verdu84:it}.

Consider the minimax problem (\ref{eqn:minimax-v2}) as a game, in which the utility function is
\begin{eqnarray*}
U_N(\underline{q}, \underline{r}) = \sup_{s \leq 0}\left[\frac{s}{N}\mathbf{E}_{p_{N, 0}}\left[g_N(\underline{Y}; \underline{q})\right] - \frac{1}{N}\log\mathbf{E}_{p_N(\cdot; \underline{r})}\left[e^{s g_N(\underline{Y}; \underline{q})}\right]\right].
\end{eqnarray*}
Both the allowable filter $\underline{q}$ and the possible operating point $\underline{r}$ are taken from $\mathcal{P} = \{\underline{x} \in [0, 1]^K: \sum_{k = 1}^K x_k = 1\}$, the space of all $K$-dimensional probability mass functions. For a given $\underline{r}$, we find that a test statistic that maximizes $U_N(\underline{q}, \underline{r})$ is the log-likelihood ratio function $g_N(\underline{y}; \underline{r}) = (1/N)\cdot\log\left[\sum_{k = 1}^K r_k p_{N, k}(\underline{y})/p_{N, 0}(\underline{y})\right]$. In conjunction with the choice of $s = -N$, this LLRT statistic leads to
\begin{eqnarray}
\max_{\underline{q}} U_N(\underline{q}, \underline{r}) = \frac{1}{N} D\left(p_{N, 0}\left\|\sum_{k = 1}^K r_k p_{N, k}\right.\right).
\end{eqnarray}
So if $\underline{r}^\ast \in \mathcal{P}$ minimizes $D\left(p_{N, 0}\left\|\sum_{k = 1}^K r_k p_{N, k}\right.\right)$, it is a least favorable operating point of the game. We now show that this least favorable operating point and its associated LLRT statistic constitute a saddle point for the game, and hence solve the minimax robustness problem, by using \cite[Thm. 2.1]{verdu84:it}. First, the set of all $K$-dimensional probability mass functions, $\mathcal{P}$, is a convex set by its definition. Second, due to the concavity of logarithmic functions and the supremum operation, the utility function $U_N(\underline{q}, \underline{r})$ is convex with respect to $\underline{r}$ on $\mathcal{P}$, for every $\underline{q}$.

It remains to be shown that $\left(g_N(\cdot; \underline{r}^\ast), \underline{r}^\ast\right)$ is a ``regular pair'', that is, if for every $\underline{r} \in \mathcal{P}$ and every sufficiently small $\beta > 0$, the perturbed distribution $\sum_{k = 1}^K \left[(1 - \beta) r_k^\ast + \beta r_k\right] p_{N, k}$ satisfies $\max_{\underline{q}} U_N(\underline{q}, (1-\beta)\underline{r}^\ast + \beta \underline{r}) - U_N(\underline{r}^\ast, (1-\beta)\underline{r}^\ast + \beta \underline{r}) = o(\beta)$ where $o(\beta)/\beta \rightarrow 0$ as $\beta \rightarrow 0$. The optimal test statistic in response to $[(1-\beta)\underline{r}^\ast + \beta \underline{r}]$ is its corresponding log-likelihood ratio
\begin{eqnarray*}
g_N(\underline{y}; (1-\beta)\underline{r}^\ast + \beta \underline{r}) = \frac{1}{N} \log\left[\frac{\sum_{k = 1}^K \left[(1 - \beta) r_k^\ast + \beta r_k\right] p_{N, k}(\underline{y})}{p_{N, 0}(\underline{y})}\right],
\end{eqnarray*}
and it follows that
\begin{eqnarray}
\max_{\underline{q}} U_N(\underline{q}, (1-\beta)\underline{r}^\ast + \beta \underline{r}) = \frac{1}{N} D\left(p_{N, 0}\left\|\sum_{k = 1}^K \left[(1 - \beta) r_k^\ast + \beta r_k\right] p_{N, k}\right.\right),
\end{eqnarray}
which behaves for $\beta \ll 1$ like \cite{zhang08:it}
\begin{eqnarray}
\label{eqn:regular-1}
&&\max_{\underline{q}} U_N(\underline{q}, (1-\beta)\underline{r}^\ast + \beta \underline{r})\nonumber\\
&=& \frac{1}{N}\left\{
D\left(p_{N, 0}\left\|\sum_{k=1}^K r_k^\ast p_{N, k}\right.\right) + \left[1 - \mathbf{E}_{p_N(\cdot; \underline{r})}\left[\frac{p_{N, 0}(\underline{Y})}{\sum_{k=1}^K r_k^\ast p_{N, k}(\underline{Y})}\right]\right] \cdot\beta
\right\} + o(\beta),
\end{eqnarray}
where the expectation is with respect to $\sum_{k = 1}^K r_k p_{N, k}$. On the other hand, $U_N(\underline{r}^\ast, (1-\beta)\underline{r}^\ast + \beta \underline{r})$ is lower bounded by setting $s = -N$,
\begin{eqnarray}
\label{eqn:regular-2}
&&U_N(\underline{r}^\ast, (1-\beta)\underline{r}^\ast + \beta \underline{r}) \nonumber\\
&\geq& \frac{1}{N} \left\{D\left(p_{N, 0}\left\|\sum_{k=1}^K r_k^\ast p_{N, k}\right.\right) - \log\mathbf{E}_{p_N(\cdot; (1-\beta)\underline{r}^\ast + \beta\underline{r})}\left[\frac{p_{N, 0}(\underline{Y})}{\sum_{k=1}^K r_k^\ast p_{N,k}(\underline{Y})}\right] \right\}\nonumber\\
&=& \frac{1}{N} \left\{D\left(p_{N, 0}\left\|\sum_{k=1}^K r_k^\ast p_{N, k}\right.\right) - \log\left[1 - \beta + \beta\cdot \mathbf{E}_{p_N(\cdot; \underline{r})}\left[\frac{p_{N, 0}(\underline{Y})}{\sum_{k=1}^K r_k^\ast p_{N, k}(\underline{Y})}\right]\right]\right\}\nonumber\\
&=& \frac{1}{N} \left\{D\left(p_{N, 0}\left\|\sum_{k=1}^K r_k^\ast p_{N, k}\right.\right) - \left[\mathbf{E}_{p_N(\cdot; \underline{r})}\left[\frac{p_{N, 0}(\underline{Y})}{\sum_{k=1}^K r_k^\ast p_{N, k}(\underline{Y})}\right] -1\right]\cdot\beta\right\}+o(\beta).
\end{eqnarray}
A direct comparison between (\ref{eqn:regular-1}) and (\ref{eqn:regular-2}) then reveals that
\begin{eqnarray*}
0 \leq \max_{\underline{q}} U_N(\underline{q}, (1-\beta)\underline{r}^\ast+\beta\underline{r}) - U_N(\underline{r}^\ast, (1-\beta)\underline{r}^\ast+\beta\underline{r}) \leq o(\beta),
\end{eqnarray*}
hence establishing the regularity of $\left(g_N(\cdot; \underline{r}^\ast), \underline{r}^\ast\right)$.

Finally, from Lemma \ref{lem:phi-ast-minimizer}, under the dominance condition (\ref{eqn:PSD-dominance-MR-LLR}), for every sufficiently large $N$, the probability distribution $p_{N, 1}$ induced by $\phi^\ast(\cdot)$ solves
\begin{eqnarray}
\min_{\underline{r} \in \mathcal{P}} \frac{1}{N} D\left(p_{N, 0}\left\|\sum_{k = 1}^K r_k p_{N, k}\right.\right).
\end{eqnarray}
So $\{p_{N, 1}\}$ and its associated LRT detector sequence achieve the minimax robust error exponent. This concludes the proof of Theorem \ref{thm:MR-LLR}.

%%%%%%%%%%%%%%%%%%%%%%%%%%%%%%%%%%%%%%%%%%%%%%%%%%%%%%%%%%%%%%%%%%%%%%%%%%%%%%%%%%%%%%%%%%%%%%%%%%%%%%%%%%%%
\section{Concluding Remarks}
\label{sec:conclusion}

Characterizing the minimax robust error exponents of hypothesis testing problems under modeling uncertainty has been a longstanding problem. This is especially the case when signal processes exhibit temporal correlation as described by PSDs, since even if the PSD uncertainty set is convex, the set of induced probability distributions generally loses the convexity property, which is usually pivotal to the existence of minimax robust detectors. In this paper, we have considered the scenario of detecting a WSS Gaussian signal processes embedded in white Gaussian noise, where the uncertainty is only with respect to the PSD of the signal process. Our treatment of the problem is based on a dominance condition, instead of the usual convexity condition, for the PSD uncertainty set. Under such a dominance condition, the minimax robust detector sequence and the resulting minimax robust error exponent are both identified. Potential future directions of interest include extending the approach in this paper to more general detection models, incorporating noise uncertainty or non-Gaussian signal/noise distributions, and exploring applications of the dominance structure among probability distributions in other problem settings.

%%%%%%%%%%%%%%%%%%%%%%%%%%%%%%%%%%%%%%%%%%%%%%%%%%%%%%%%%%%%%%%%%%%%%%%%%%%%%%%%%%%%%%%%%%%%%%%%%%%%%%%%%%%%
\bibliographystyle{ieee}
\bibliography{./minimax_detect}

\end{document}